\documentclass[acmsmall]{acmart}

\AtBeginDocument{%
  }

\acmSubmissionID{fse24main-p761-p}

\setcopyright{rightsretained}
\acmDOI{10.1145/3643766}
\acmYear{2024}
\copyrightyear{2024}
\acmSubmissionID{fse24main-p761-p}
\acmJournal{PACMSE}
\acmVolume{1}
\acmNumber{FSE}
\acmArticle{40}
\acmMonth{7}
\received{2023-09-29}
\received[accepted]{2024-01-23}

\newboolean{showcomments}
\setboolean{showcomments}{false}

\newboolean{showboxes}
\setboolean{showboxes}{true}

\usepackage[utf8]{inputenc}
\usepackage{url}
\usepackage{colortbl}
\usepackage{balance}
\usepackage{xspace}
\usepackage{graphicx}
\usepackage{verbatim}
\usepackage{bold-extra}
\usepackage{paralist}
\usepackage{multirow}
\usepackage{listings}
\usepackage{boxedminipage}
\usepackage{soul}
\usepackage{enumitem}
\usepackage[normalem]{ulem}
\setlist{nolistsep,leftmargin=.5cm}
\usepackage{booktabs}
\usepackage{tabularx}
\usepackage{svg}
\usepackage{calc}
\usepackage{float}
\usepackage{multirow}
\usepackage[normalem]{ulem}
\useunder{\uline}{\ul}{}
\usepackage[most]{tcolorbox}
\usepackage{caption}
\usepackage{microtype} 
\usepackage{algorithmic}
\usepackage{graphicx}
\usepackage{textcomp}
\usepackage{xcolor}
\usepackage{hyperref}
\usepackage{ifthen}
\usepackage{wrapfig}
\usepackage{subcaption}
\usepackage{cleveref} 
\usepackage{adjustbox}
\usepackage{totcount}

\ifthenelse{\boolean{showcomments}}
{\newcommand{\nb}[2]{
		\fbox{\bfseries\sffamily\scriptsize#1}
		{\sf\small$\blacktriangleright$\textit{#2}$\blacktriangleleft$}
	}
	
}
{\newcommand{\nb}[2]{}
	
}

\newcommand{\eg}{\textit{e.g.},\xspace}
\newcommand{\etc}{\textit{etc.}\xspace}
\newcommand{\etal}{\textit{et al.}\xspace}

\newcommand{\ellipsis}{\textit{…}\xspace}
\newcommand{\figone}{Figure 1\xspace}
\newcommand{\figtwo}{Figure 2\xspace}

\newcommand\rev[1]{{{#1}}}

\newcounter{findingcounter}
\newtotcounter{totalfindings}
\ifthenelse{\boolean{showboxes}}
{
    \newcommand{\finding}[1]{%
      \refstepcounter{findingcounter}
      \begin{tcolorbox}[boxsep=1pt,left=2pt,right=2pt,top=1pt,bottom=1pt]%
      \small
      \centering
      \textbf{Finding \arabic{findingcounter}:} #1
      \end{tcolorbox}%
      \addtocounter{totalfindings}{1}
    }
    
  }
{
    \newcommand{\finding}[1]{}
}

\begin{abstract}

Most modern software products incorporate open source components, which requires compliance with each component's licenses. As noncompliance can lead to significant repercussions, organizations often seek advice from legal practitioners to maintain license compliance, address licensing issues, and manage the risks of noncompliance. While legal practitioners play a critical role in the process, little is known in the software engineering community about their experiences within the open source license compliance ecosystem. To fill this knowledge gap, a joint team of software engineering and legal researchers designed and conducted a survey with 30 legal practitioners and related occupations and then held 16 follow-up interviews.  We identified different aspects of OSS license compliance from the perspective
of legal practitioners, resulting in \total{totalfindings} key findings in three main areas of interest: the general ecosystem of compliance, the specific compliance practices of legal practitioners, and the challenges that legal practitioners face. We discuss the implications of our findings.
\end{abstract}

\begin{document}

\title{``The Law Doesn't Work Like a Computer'': Exploring\\ Software Licensing Issues Faced by Legal Practitioners}

\author{Nathan Wintersgill}
\orcid{0009-0006-2123-7412}
\affiliation{%
  \institution{William \& Mary}
  \city{Williamsburg, VA}
  \country{USA}
}
\email{njwintersgill@wm.edu}

\author{Trevor Stalnaker}
\orcid{0009-0005-6000-4227}
\affiliation{%
  \institution{William \& Mary}
  \city{Williamsburg, VA}
  \country{USA}
}
\email{twstalnaker@wm.edu}
\author{Laura A. Heymann}
\orcid{0000-0001-9258-2105}
\affiliation{%
  \institution{William \& Mary}
  \city{Williamsburg, VA}
  \country{USA}
}
\email{laheym@wm.edu}

\author{Oscar Chaparro}
\orcid{0000-0003-2838-685X}
\affiliation{%
  \institution{William \& Mary}
  \city{Williamsburg, VA}
  \country{USA}
}
\email{oscarch@wm.edu}

\author{Denys Poshyvanyk}
\orcid{0000-0002-5626-7586}
\affiliation{%
  \institution{William \& Mary}
  \city{Williamsburg, VA}
  \country{USA}
}
\email{denys@cs.wm.edu}

\renewcommand{\shortauthors}{Nathan Wintersgill, Trevor Stalnaker, \\Laura A. Heymann, Oscar Chaparro, and Denys Poshyvanyk}



\maketitle

\sloppy

\section{Introduction} %

Over more than two decades, the open source software (OSS) community has propelled the software industry forward, creating a dynamic supply chain where software systems are often built by integrating existing OSS components, such as libraries and frameworks~\cite{synopsys_report, ghosh2007economic, blind2023estimating, 8728094}. This reuse of components allows developers to forgo reinventing the wheel by freely utilizing existing solutions to common problems and thus accomplish their tasks more productively. To promote open collaboration and support the software supply chain, OSS components are typically released under one or more OSS licenses, which define the terms governing the components’ distribution, modification, and reuse.
As software is protected by copyright and patent laws~\cite{copyright_software}, software projects integrating such components must comply with the terms of any applicable licenses, and failing to do so can result in significant 
legal~\cite{vizio_1, ai_copyright}, reputational~\cite{johndeere, gunningham2004social}, and financial consequences~\cite{cost1, cost2} for developers and organizations.

Although ensuring software license compliance is a critical process for developers and organizations, 
the process is often challenging
~\cite{german2009license,gangadharan2012managing,almeida2019investigating,vendome2018distribute}. 
Software systems often integrate hundreds or even thousands of OSS components, distributed under one or more licenses~\cite{8728094}. 
These licenses can conflict with one another, as there are hundreds of OSS licenses with different (in)compatibility levels~\cite{spdx_licenses,osi}.
Additionally, licenses are written in legal terms that can be subject to different interpretations,  %
and developers often struggle to accurately apply these terms~\cite{almeida2019investigating,gangadharan2012managing,vendome2018distribute}.
\looseness=-1

Given the risks of license noncompliance, organizations often seek legal advice from in-house or outside counsel on license compliance issues, guidance on addressing such issues, and strategies for maintaining compliance. While legal practitioners play a critical role in the process of license compliance, little is known in the software engineering (SE) community about their perception and experiences within the OSS license compliance ecosystem, including their methodologies for assisting organizations during license compliance and the challenges they face in this process. Prior studies investigated OSS license compliance issues and their impact mostly from a developer/user perspective ~\cite{almeida2019investigating,vendome2017license,vendome2018distribute,wu2017analysis,papoutsoglou2022analysis}, rather than from a legal perspective.
\looseness=-1

To fill this knowledge gap in the SE community, we conducted a qualitative study that examined the experiences of a group of legal practitioners specializing in OSS license compliance in the U.S. The study, conducted by a joint team of SE and legal researchers, surveyed 30 legal professionals and compliance experts, who answered an online survey that probed into various aspects of their work and past experiences. We subsequently conducted interviews with 16 of the respondents to delve deeper into their experience within the OSS license compliance ecosystem. 

The study yielded \total{totalfindings} main findings about OSS license compliance as viewed through the lens of legal practitioners. These findings encompass various dimensions, including license selection, creation, proliferation, evolution, and interpretation; license enforcement; how licensing issues are resolved; risk management strategies; tool usage for license compliance; and prevention strategies, including training, education, and communication with developers. We qualitatively analyze these findings  
and discuss opportunities for improvement as well as avenues for future research.

In summary, our main contributions are: (1) a rigorous assessment of the current state of the OSS license compliance ecosystem from the perspectives of 30 legal practitioners and compliance experts; (2) an in-depth analysis of how these individuals perform license compliance tasks and the associated challenges they face; and (3) a thorough discussion of the findings and their implications, highlighting avenues for future work.
We provide a replication package %
for verifiability ~\cite{anonymous_repo}.

\section{Background}

Software, including OSS, is protected by copyright law in the United States as a general matter~\cite{copyright_software}. %
The owner of copyright in a work has several rights under U.S. copyright law, including the right to reproduce the work, to create derivative works, and to distribute the work, as well as the right to authorize others to engage in these activities~\cite{USCode17sect107}. %
Although the copyright in a work can be sold or transferred, a copyright license is the mechanism by which a copyright owner retains copyright in the work while authorizing another party to use the work in ways that would otherwise constitute infringement, sometimes subject to stated conditions~\cite{meeker2017open}. %

The developer of OSS typically chooses an existing OSS license under which to issue their work rather than creating a new license. At present there are 117 licenses recognized by OSI~\cite{osi_approved} and 606 by SPDX~\cite{spdx_licenses}, with a handful, such as MIT, GPL, and Apache, being the most prevalent~\cite{prevalent}. OSS licenses generally fall into one of two categories: permissive and restrictive (also called copyleft). Permissive licenses (such as MIT~\cite{mit} and BSD~\cite{bsd}) typically impose few restrictions and requirements on those using a piece of software (\eg the requirement to provide notice files containing OSS component attribution or licenses), whereas copyleft licenses (GPL~\cite{gpl_license}, LGPL~\cite{lgpl_license}, \etc) typically require that the full source code of the resulting work be made available and that the project is offered under the same license (leading some to characterize such licenses as ``viral'').

Because software licenses typically contain conditions on use, entities that use OSS components in their own development projects must ensure that such use complies with the requirements of the license to avoid a claim of copyright infringement. %
Prior studies have shown that this task can be challenging for developers to complete on their own ~\cite{vendome2018distribute}.
Companies often hire legal counsel (in-house or outside) to assist with compliance tasks and to navigate the myriad of potential licensing issues~\cite{vendome2018distribute}. These professionals provide their clients with guidance on license interpretation, create educational resources for developers, and help clients manage the risks associated with using OSS.

OSS licensing can involve several areas of law beyond copyright law, such as patent law, trademark law, cybersecurity, and privacy law. We focus in this study on U.S. copyright law and legal practitioners located in the U.S., although their work may involve jurisdictions around the world.
\looseness=-1

\vspace{-.1cm}
\section{Study Design}

This study, which involves a collaboration between SE and law researchers, aims to analyze OSS license compliance issues from the legal practitioner perspective through surveys and follow-up interviews. We address the following research questions (RQs):

\begin{enumerate}[label=\textbf{RQ$_\arabic*$:}, ref=\textbf{RQ$_\arabic*$}, wide, labelindent=5pt]\setlength{\itemsep}{0.2em}
    \item \label{rq:1}{\textit{What is the ecosystem of license compliance experienced by legal practitioners?}} 
    \item \label{rq:2}{\textit{How do legal practitioners perform license compliance in this ecosystem?}}
    \item \label{rq:3}{\textit{What challenges do legal practitioners face during license compliance?}}
\end{enumerate}

In \textbf{RQ1}, we investigate how OSS licensing compliance \textit{generally} works from the experience of legal practitioners, including how OSS licenses are selected and used, the kinds of violations that occur, how licenses are enforced, and how disputes are resolved. \textbf{RQ2} aims to investigate \textit{specific} compliance activities performed by practitioners, including when compliance is performed, who is involved in the process, how risk is managed, how education/training is performed, and how tooling is used. \textbf{RQ3} investigates \textit{challenges} faced by legal practitioners during the process.

The study combined an online survey and follow-up interviews conducted with legal practitioners and others specializing in OSS licensing within the U.S. We used various strategies to find potential participants and used an open-coding methodology to analyze survey and interview responses. This section details the research methodology (see \figone), \Cref{sec:results} presents the study results and their analysis, and \Cref{sec:implications} discusses the implications of our findings. The methodology was approved by the ethical board of our university.

\Crefformat{subfigure}{#2Figure~#1#3}

\begin{figure}
    \begin{subfigure}{0.64\textwidth} %
        \centering
        \includegraphics[width=\linewidth]{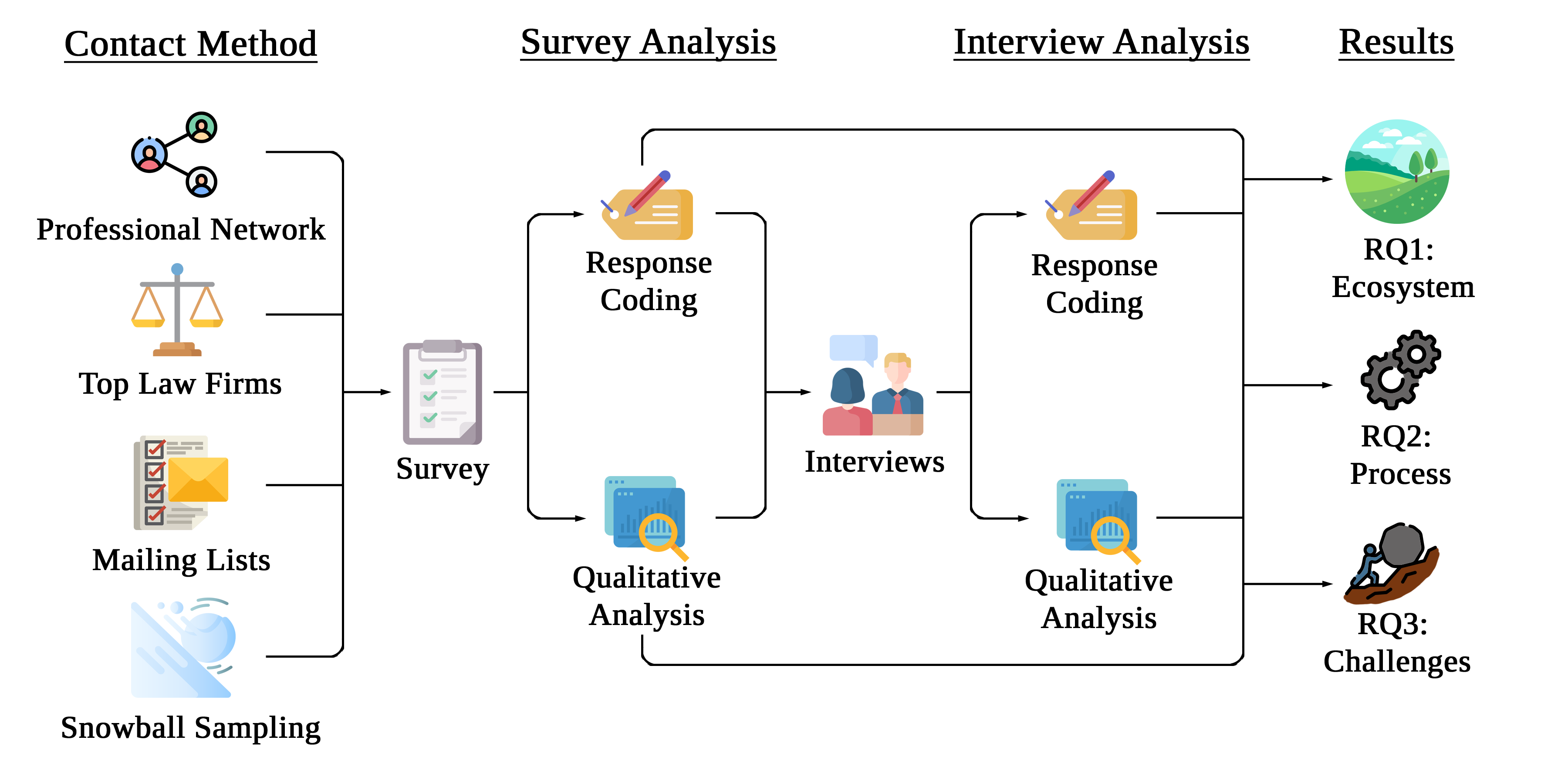}
        \label{fig:methodology}
        \caption*{Fig. 1. Research methodology \\ (image credits at \cite{anonymous_repo})}
    \end{subfigure} \hspace{0.03\textwidth}%
    \begin{subfigure}{0.29\textwidth} %
        \centering
        \includegraphics[width=\linewidth]{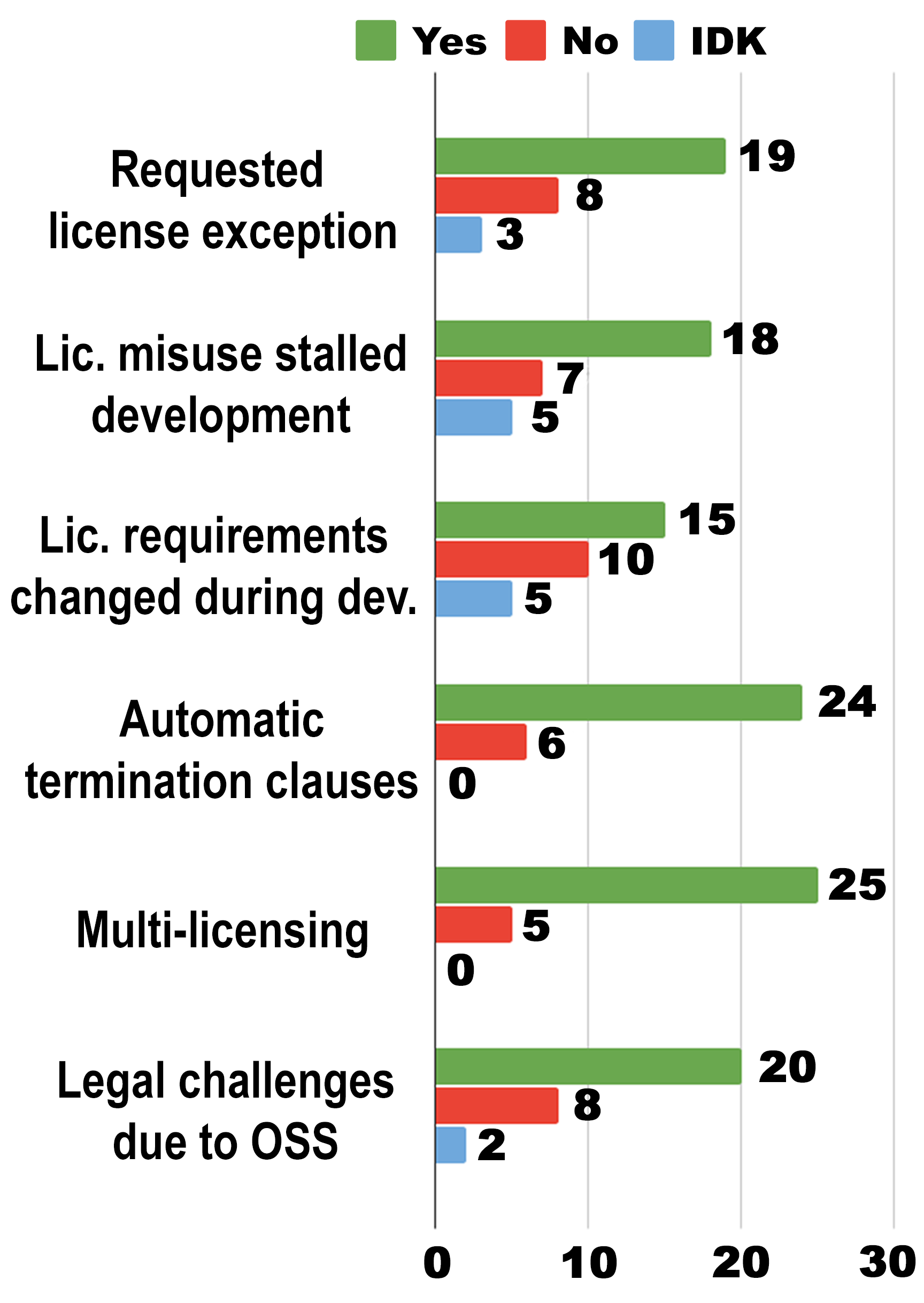}
        \caption*{Fig. 2. Practitioner experiences}
        \label{fig:experience}
    \end{subfigure}
    \vspace{-5mm}
\end{figure}

\subsection{Survey Design and Participant Identification}

In designing our survey, we followed general guidelines for survey design \cite{survey} as well as SE-specific guidelines \cite{DBLP:journals/sigsoft/PfleegerK01,DBLP:journals/sigsoft/KitchenhamP02,DBLP:journals/sigsoft/KitchenhamP02a,DBLP:journals/sigsoft/KitchenhamP02b,DBLP:journals/sigsoft/KitchenhamP02c,DBLP:journals/sigsoft/KitchenhamP03}. The survey questionnaire went through multiple iterations with reviews and modifications from four SE and three law researchers, making sure that
questions were written clearly and concisely to avoid confusion and bias.
 \rev{The survey included questions that sought to elicit information related to each of the RQs noted above, including questions relating to existing compliance practices, the use of tooling or other automation to perform compliance, and information needs and another challenges.} %
Respondents were asked to self-identify as in-house or outside counsel or as other roles. 
The full survey text and a brief description are found in our replication package~\cite{anonymous_repo}.

Our survey targeted legal practitioners with OSS experience. Since the OSS legal community is relatively small, compared to the OSS developer community~\cite{stalnaker2024boms}, %
we adopted different strategies to disseminate our survey, trying to reach as many people as possible. First, we posted the survey in Cyberprof, a listserv dedicated to Internet law educators. %
Second, we compiled a list of top legal firms in the United States, ranked by gross revenue, from the Am Law 100 list~\cite{AMLaw}. Potential participants were identified by manually visiting the firm websites and employee directories and collecting public contact information of people with relevant expertise, resulting in 724 email addresses. %
Third, we reached out to colleagues from our professional (SE and legal) networks. Fourth, we applied snowball sampling by asking potential participants to disseminate the survey to their networks, including private mailing lists. During interviews (described in \Cref{sec:interviews}), we also asked participants for people with whom we should share the survey.

\vspace{-.2cm}
\subsection{Survey Response Collection and Analysis}

Responses from survey participants were collected using Qualtrics~\cite{qualtrics}; the survey was kept open for six weeks.
\Cref{tab:demographics} shows that 32 complete survey responses were obtained. Of these, 30 were valid; we discarded two invalid responses that answered the questions with spurious answers. %

Twenty respondents self-identified as active in-house/outside legal counsel and ten as other roles.  All have worked professionally in the area of OSS licensing, ten with more than 20 years of experience.
Most participants were trained as lawyers (although not all were actively practicing); some respondents were not trained as lawyers but were actively engaged in OSS compliance efforts.
All respondents came from our professional network, mailing lists, and snowball sampling. Notably, no lawyers from the top firms completed our survey, which illustrates the difficulties of surveying the OSS legal population.
\looseness=-1

We qualitatively analyzed the survey responses by applying a coding approach, in line with~\cite{spencer2009card}. Two SE researchers and a law student (hereon ``annotators'' %
) performed \textit{open coding}, independently assigning one or more codes to each response using a shared spreadsheet. Each annotator independently coded all 32 complete responses, adding new codes to the spreadsheet when appropriate. Once open coding was completed, the annotators convened to settle disagreements and consolidate the set of codes. Our replication package~\cite{anonymous_repo} contains the final codes and definitions.
We did not base our analysis on inter-rater agreements since multiple codes could be assigned to each response and no list of codes existed prior to the start of coding. 
Where there was a high level of dispute or the response was ambiguous, an additional law researcher offered input. Where appropriate, results were analyzed by leveraging descriptive statistics on the assigned codes.

\vspace{-.2cm}
\subsection{Interview Design and Analysis}
\label{sec:interviews}
We conducted 16 semi-structured interviews over Zoom with survey participants who %
expressed a willingness for a follow-up conversation and whose survey responses %
warranted further investigation. (Two additional individuals were contacted but did not respond to our interview request.) Interviewees were selected to represent a range of backgrounds and professional experience. The interviews were designed to obtain deeper knowledge about respondent experiences and responses. %
Each interview lasted 30 to 60 minutes and was recorded to facilitate transcription and subsequent analysis. Both SE and law researchers prepared for and were present at each interview.  %

Each interview (recording and transcript) was independently reviewed by one SE researcher and one law researcher. %
A shared text document was used to aggregate findings from across interviews by assigning topic labels to all relevant portions of each response.
To validate the accuracy and completeness of the final document, all reviewers, three in total, met to discuss and come to a consensus on the framing and application of topic labels.

To preserve the anonymity of our interviewees, we assigned each interviewee an identifier to attribute the source of quotations, as noted in Table 2. 
Respondents R1, R3, R4, R5, R6, R7, R15, and R16 have licensing experience as lawyers in private practice in small or mid-sized firms; respondent R8 has licensing experience as a lawyer in private practice in a large firm. Respondents R2, R12, and R13 have licensing experience as current or recent in-house counsel for major technology companies. (Some of the lawyers in private practice also have in-house experience and vice versa.) Respondent R10 has licensing experience as in-house counsel to an organization focused on free and open source software, and respondents R9 and R14 have licensing experience as nonlawyer employees of organizations focused on OSS license compliance. Respondent R11 has licensing experience as an open source program office director at a major technology company. We have omitted filler words (\eg ``like,'' ``you know'') in quotations throughout and have indicated edits made for clarity with brackets and ellipses.
\looseness=-1

\begin{table}[]
\caption{Number of survey responses and respondent demographics}
\label{tab:demographics}
\begin{tabular}{|lc|lc}
\cline{1-4}
\multicolumn{2}{|c|}{\textbf{Survey}}                                                & \multicolumn{2}{c|}{\textbf{Other roles}}                             \\ \cline{1-4}
Total responses                                                 & 61                 & Professor                          & \multicolumn{1}{c|}{3}           \\
Complete responses                                              & 32                 & Compliance Director / Expert       & \multicolumn{1}{c|}{3}           \\ \cline{1-2}
\multicolumn{2}{|c|}{\textbf{Valid responses by current role}}                       & Executive Director                 & \multicolumn{1}{c|}{1}           \\ \cline{1-2}
In-house counsel                                                & 12                 & Enforcement Professional           & \multicolumn{1}{c|}{1}           \\
Outside counsel                                                 & 8                  & Open Source Manager / Director     & \multicolumn{1}{c|}{2}           \\  
Other roles                                                     & 10                 &                                    & \multicolumn{1}{l|}{}            \\ \cline{1-4}
\textbf{Total}                                                  & \textbf{30}        & \multicolumn{1}{l}{\textbf{Total}} & \multicolumn{1}{l|}{\textbf{10}} \\ \cline{1-4}
\end{tabular}%
\end{table}

\begin{table}[]
\caption{Interview participants by most recent source of licensing experience}
\label{tab:my-table}
\begin{tabular}{|l|l|}
\hline
\multicolumn{1}{|c|}{\textbf{Role (as described in interview)}} & \multicolumn{1}{c|}{\textbf{Interview participants}} \\ \hline
Private practice & R1, R3, R4, R5, R6, R7, R8, R15, R16 \\
In-house counsel & R2, R10, R12, R13 \\
Employee of compliance organization & R9, R14 \\
OSPO director & R11 \\ \hline
\end{tabular}%
\end{table}

\section{Results}
\label{sec:results}

\subsection{RQ1: The Ecosystem of OSS License Compliance Experienced by Legal Practitioners}
\renewcommand{\ref}[1]{RQ\,\textup{\ref{#1}}}%

We discuss how OSS license compliance is generally experienced by legal practitioners. In particular, we discuss how licenses are selected and used, the nature of license violations, how licenses are enforced to address violations, and how disputes are resolved.

\subsubsection{\textbf{License Selection and Proliferation}}
Legal practitioners encourage OSS teams to use common, pre-drafted licenses (\eg those approved by OSI), as it reduces the effort required to understand a license's terms\rev{; as Respondent R10 noted,} ``One of the main advantages of open-source software\ellipsis is [that] they're templates\ellipsis. These aren't negotiated every single time, so lawyering to some extent really comes down to the most basic skill of cutting and pasting.''  There may not, however, be an existing OSS license that exactly fits a licensor's needs. This could prompt OSS teams  
to create new, bespoke licenses, thus causing license proliferation~\cite{prolif}. \rev{Respondent R2} characterized this as undesirable because it increases license compliance efforts: 
``[W]hen someone takes one of the licenses and slightly modifies it, or tries to write their own license, then even when the license seems fairly innocuous, because it's an unknown, it requires more vetting by the powers that be.'' %
Interviewees seemed to agree, consistent with the views of OSS organizations~\cite{FSF_home,osi,github}, that writing new licenses is usually not necessary and even harmful\rev{.} \rev{Respondent R4 stated,} ``I think the majority of folks have tended to say, `We want to have as few new licenses as possible. The more licenses there are, the more compatibility issues there are.'\thinspace''  The consensus appears to be that license proliferation has waned in recent years; as \rev{Respondent R5} noted, ``[T]hat was a significant problem in the mid-2000s, but I don't see it happening anymore\ellipsis. [E]verybody understands that one of the biggest benefits of open source is a known license that you understand the characteristics of, and you don't have to read it and figure out what it says.'' %

\vspace{-3pt}
\finding{Legal practitioners encourage the use of off-the-shelf OSS licenses. License proliferation\\
can obviate the main benefits of off-the-shelf OSS licenses but is reportedly no longer a common practice.}

\subsubsection{\textbf{License Exceptions}} The off-the-shelf nature of OSS licenses also has implications for licensees, who might desire to engage in uses not covered by the applicable license. In these cases, legal practitioners can help their clients obtain license exceptions (or additional permissions), which allow clients to operate under different terms from the original ones provided in the license. This practice is not uncommon: 19 survey respondents indicated that they had experience negotiating or requesting license exceptions (see \figtwo). Experiences regarding such requests varied. Six respondents noted that exceptions were difficult to obtain, while five cited them as easy to acquire.

Respondents noted two challenge categories that may arise while seeking or managing license exceptions. One is the difficulty of contacting all relevant developers for a component. A developer may have died, or permission may need to be secured from multiple developers, depending on the number of dependencies. As  \rev{Respondent R15} put it, ``If you look at GitHub repositories, you would be amazed at how many stale links are on there, or how many people have moved or transitioned\ellipsis.'' The other is the informal nature of documenting exceptions, which \rev{Respondent R4} described as either an internal note or a statement in the repository of a public project: ``[T]he technical person goes and contacts the author of project X and says, \ellipsis `Could you send us an email just confirming you're okay with x, y, and z?'\ellipsis And then the client goes forward and just relies on that \rev{[response]}.'' Informal documentation can lead to difficulties when performing license compatibility checks or addressing potential license violations.

\finding{Obtaining license exceptions is not uncommon but can be challenging.  It can be difficult \\or impossible to even start the process if project authors cannot be identified or reached.} %

\subsubsection{\textbf{Licensing Changes and Multi-licensing}}
Project maintainers may decide to change the license of their products (\eg from OSS to a proprietary/business license \cite{terraform_relicensed} or between two OSS licenses \cite{vlc_relicensed}) for a variety of reasons, such as changes in business strategy~\cite{di2010exploratory, vendome2017license}. These license changes can impact a project's dependents. In some cases, the new license may not align with a dependent organization's goals (\eg a closed-source project using software that requires source code disclosure). This leaves developers with two options\rev{, as Respondent R4 described}: ``[E]ither we switch to the new license and have to figure out whether we can [meet the new terms] \rev{and what the implications are}, or we stay on the old license [and forgo] security updates from the mainstream branch\ellipsis.'' Another possibility is to wait for another member of the community to fork the original project and maintain it \rev{separately} \cite{terraform_fork,mariadb}, such as when ``[p]eople forked [MySQL], and now there's a competing product called MariaDB, which is GPL licensed'' (R10). %

 Interviewees \rev{also} discussed the practice of multi-licensing, which \rev{encompassed} various situations, including software released under more than one license; software containing dependencies or files under different licenses; and the offering of a free restrictive license with a %
paid option for %
a more permissive license (``[Y]ou release under AGPL, and then some people can't use code under AGPL, so they want to buy exceptions from you.'' \rev{(R6)}). \rev{A} software project can be multi-licensed unintentionally if it is distributed through multiple channels under different licenses\rev{, although it is unclear how often this happens. As} Respondent R15 described it, ``On site A, it says that this is licensed pursuant to the LGPL [but on] site B, it says it's licensed pursuant to the MIT license\ellipsis. \rev{[T]hat ambiguity becomes a huge risk for my clients}\ellipsis.'' %

\finding{License changes can happen for a variety of reasons, but they can be difficult to\\ negotiate and can impact dependent projects. Proprietary relicensing can result in \\project forking. The term \textit{multi-licensing} has different meanings depending on context.}

\subsubsection{\textbf{License Violations}} 
The goal of license compliance is, of course, to avoid violations.
In the 30 survey responses, several key themes emerged pertaining specifically to license violations, including conflicting license terms (6), missing source code required by certain licenses to be distributed (4), and software that is missing license information (3) or copyright information (1). We further explored such violations in follow-up interviews. Three interviewees indicated that issues tended to arise from software developers copying or using non-compliant code without considering its licensing information. \rev{As Respondent R11 described,} ``[P]eople find stuff on GitHub, and they don't pay attention to the fact that there's no license in the repo and no license information in any of the source files. They just figure since it's up on GitHub, it's intended to be shared\ellipsis. It happens weekly.''
\looseness=-1

Multiple respondents indicated that they perceived license violations to be very common: ``There is rampant non-compliance in the field,'' Respondent R6 indicated in their survey response; Respondent R7 said, ``[In] open source licensing, a lot of violations are ignored. People just kind of don't care.''  Some respondents characterized noncompliance as knowing and intentional, with compliance pursued only in response to complaints. Others indicated that being a ``good citizen'' was particularly important in open source communities and suggested that it was very difficult to prevent every violation. 
For example, when asked how organizations they have worked with ensure license compliance, \rev{the} survey respondent \rev{mentioned above} indicated that ```[e]nsured' is impossible.'' %
Instead, compliance becomes a risk management activity in which risk might never be fully eliminated but can be mitigated, as \rev{Respondent R7} indicated: ``[Tools] together with people who are knowledgeable on this stuff and processes that are designed to resolve problems can help organizations \ellipsis put together a process that is reasonably risk-mitigating\ellipsis.'' %
\looseness=-1

Participants indicated that license violations can have important negative impacts not only on licensors and consumers (who might not receive source code to which they are entitled) but also on developers and their companies. As shown in \figtwo,  18 survey respondents indicated that organizations or clients they worked with had their development or release process stalled due to improper usage of licensed software. Eight of these indicated that licensing issues required software re-engineering, and six indicated that such issues led to delays in shipping software.
\looseness=-1

\finding{Non-compliance is frequent in the OSS ecosystem, resulting, in part, from 
difficulties in maintaining compliance. License compliance violations impact non-compliant parties, software users, and the ecosystem at large. The development/release process can ultimately be stalled due to licensing issues.}

\subsubsection{\textbf{License Enforcement and Dispute Resolution}} 
When license enforcement occurs, our results show that it is typically performed by open source community organizations rather than directly by the parties whose licenses have been violated\rev{, as Respondent R6 indicated}: ``[C]ommunity compliance actions [are] by far the most common\ellipsis. [Organizations like the Software Freedom Conservancy] are kind of like pro bono organizations.'' Regardless of the enforcing entity, \rev{the fact that resources are limited means that enforcement decisions can be} based on the perceived importance of a violation\rev{, as Respondent R13 noted}: ``The community focuses their limited resources on where they can make the biggest impact, which means that there's different treatment depending on who you are.'' %
\looseness=-1

License compliance disputes can be resolved without litigation. For example, enforcement organizations might extend the ``cure period'' in a license (the grace period during which a non-compliant licensee can come back into compliance without risking license termination). \rev{Respondent R9} cited a case in which a license had a cure period of 30 days, but legal action was not brought until two years later.  Survey respondents were asked how the organizations they have worked with resolved legal challenges related to the usage of open-source software; while litigation (7) and settlements (6) were mentioned by several respondents, a number of respondents also indicated that they resolved the matter informally with the other party (4), or by rewriting/removing non-compliant code (3) or otherwise bringing their software into compliance (2).

Respondents indicated that litigation is rare, and several interviewees stated that it is often viewed unfavorably in the OSS community. Respondent R7 pointed out that ``[l]itigation is an expensive process, and oftentimes results in decisions that may not achieve the result that one desires -- on either side,'' while Respondent R5 noted that ``there's a great deal of antagonism in the open-source community over whether or not lawsuits are appropriate at all.'' (Respondent R5 also noted, however, that ``strategic lawsuits can be very effective -- you just need that one lawsuit to make your point.'') Thus, if a license issue is brought to a developer's attention, one option is to simply apologize and correct the issue, either by coming into compliance or rewriting the code, \rev{as Respondent R12 noted}: ``I think  normally just falling on your sword and saying, `Look, we made a mistake here,' and then working out some sort of solution is usually what happens.'' %
Additionally, public shaming -- issuing statements accusing an organization of flouting license compliance -- can pressure organizations into complying to avoid reputational damage. As \rev{Respondent R6} put it, ``Public shaming is a very effective strategy\ellipsis. %
[T]echnology companies in particular are in hot competition for engineering resources all the time\ellipsis. And if you get known as an open source scofflaw, you have trouble recruiting people.'' %
\looseness=-1

\finding{Enforcement is typically done by the community rather than by licensors. Due to limited resources, enforcement is unevenly carried out, with a perceived overall lack of enforcement. Litigation is rare, with other enforcement methods (\eg public shaming or informal resolution) used more frequently.}

\subsection{RQ2: The Process of OSS License Compliance from a Legal Perspective}
\label{sec:compliance}

We discuss different facets of the OSS license compliance process experienced by legal practitioners.
\looseness=-1

\subsubsection{\textbf{When Compliance is Done}}

Respondents indicated that compliance tasks can be done reactively or proactively. %
\rev{In Respondent R9's experience}, many companies ``approach license compliance as something that you do at the end to make sure that you're able to combine all of these things in a way that meets all of the licenses.''  However, \rev{Respondent R9 noted,} when this approach is taken for ``whatever jumble of code you happen to have [at] the end, there's a good chance there will be problems, and they will be difficult to deal with.'' Legal teams are acutely aware of the problem; according to \rev{Respondent R7}, ``[S]oftware folks are not real happy when [a component needs to be replaced] because you can't just immediately rip stuff out, rewrite it, and test it when you've got a ship date that's  coming up very soon.'' The impact is heightened when copyleft licenses are involved; \rev{as Respondent R12 noted, if a compliance review happens only at the end of a project,} ``and you find out you've got something that's GPL-related, now you've got obligations to release the rest of the code along with the GPL stuff under those licenses. And you either have to do that or you have to pull it and find a replacement for the code that you have.'' %
\looseness=-1

Alternatively, compliance can be done throughout the development cycle.  This approach comes with distinct advantages. \rev{Respondent R9} mentioned that if \rev{teams} ``do some assessments as you're going along in building the project, then I think it is much easier to both make the determination and then to comply at that point, \rev{or to make the small changes you may need to make to comply with the license}.''   Incremental scans and thoughtful compliance processes, \rev{as Respondent R4 noted,} ``help steer off those issues early on so that the team can say, `Okay, this will be an issue if we use this dependency. Let's find a different one that's under a different license that's going to work better.'\thinspace'' Nevertheless, best practices typically still involve a final scan; \rev{as Respondent R12 stated,}  ``[W]hen you have a lot of hands in the pot, \rev{\ellipsis I think, if you're doing it right, you have to assume that somebody} might just slip something in there without having checked it in.'' %

\finding{Compliance is less difficult and more effective if it occurs \\throughout the development process rather than solely at the end.}

\subsubsection{\textbf{Actors/Roles}}
\label{ssec:actors}

Multiple actors/roles are involved in the compliance process, notably developers and lawyers.
But who within an organization is responsible for carrying out compliance tasks?  
According to some participants, while the impacts of non-compliance are felt organization-wide, the ultimate responsibility should be on the developers, since they are the closest to the code base.  The role of legal teams, \rev{said Respondent R11,} is to ``provide the tools that can help make compliance easier, the training that can help them understand why it is important, best practices or playbooks\ellipsis. But ultimately it's the developer's responsibility. They're the ones closest to the code. They're the ones touching it.'' %
\rev{Respondent R2 was in accord:} ``I'm not searching through the code base as a lawyer on a daily basis and I, to some extent, have to rely on the developers to tell us [what's in the project].''
\looseness=-1

To alleviate the burden on developers and assist them in compliance tasks, some mid- to large-size organizations have Open Source Program Offices (OSPOs). Companies created OSPOs, \rev{in Respondent R7's view,} ``because they started to realize they needed to coordinate the efforts that they were doing internally around open source, and they needed to understand how the appropriate ways to interact with open source communities were. The mindset of proprietary software didn't work when you were working within these communities.''  Most OSPOs, \rev{according to Respondent R7,} include an ``attorney that is a liaison with or part of that organization, who kind of gets the job of understanding the basics of open source licensing.'' %

\rev{Respondent R11} described how their company's OSPO operates: ``\rev{[W]e've got a number of practitioners who are open source savvy or open source experts.} We've got some tooling that's used to scan software that we've developed before it goes out. And we educate developers on how they should go about doing what we call IP planning for their projects, which is basically understanding what the proposed distribution model is going to be. You kind of have to understand that at the outset to figure out what the guardrails are going to be. And then identifying any third-party software that they're planning to be using, making sure that it's under licenses that are going to allow us to use it in the way that we want to be using it. And then just sort of incremental checkpoints along the way.'' In addition to these functions, an OSPO can help ensure that a company has, \rev{in Respondent R7's view,} ``the pulse of the open source movement,'' so that it is ``doing things in a way that [is] productive'' and does not ``cause backlash within this community that [it's] trying to work with.'' %
\looseness=-1

\vspace{-0.1cm}
\finding{Compliance is a team effort among developers, lawyers, and other roles. \\However, compliance is seen as primarily a developer's responsibility. OSPOs are a way \\that larger organizations can manage license compliance and assist developers in the process.}

\subsubsection{\textbf{Provenance and Recordkeeping}}

One of the major challenges in open source — something ``software attorneys obsess about'' \rev{(R10)} — is determining what is in a project, where it originated, and what licenses attach to each component. Twenty-four of 30 survey participants indicated that provenance was essential information needed for compliance. \rev{Respondent R16, a lawyer,} stated that if this information is not tracked at the time the component is copied, it can be ``very difficult to go back and forensically reconstruct how they got [to] where they are now with open source technologies.''  \rev{Respondent R1, also a lawyer,} noted that issues can arise when developers ``know that something is open source, but they won't know the nature of the license under which they grabbed it because [\rev{the webpage has since been updated}]\ellipsis. [W]e can't get the original one unless we go back in the Wayback Machine, which is hit or miss.''   \rev{Respondent 14}, a non-lawyer with expertise in compliance, had a contrary view: ``I can tell you what's in your products within minutes\ellipsis. It's not like folks have spent \ellipsis time hiding it, and it's not like their upstreams are covering up the fact that there's Linux in there or any other copyleft program.''  (The different views may result from whether the task is to track all software components in a build or to locate a specific license violation associated with a single component.)

One way that developers can take on this responsibility is by keeping detailed and accurate records of where code and dependencies are pulled from. The need for recordkeeping becomes acute in the situation where, \rev{as Respondent R16 noted,} ``it's years after the fact, contractors have left, [and] there's been a brain drain from the organization.'' %
The size of an organization can have an impact on the importance of recordkeeping, \rev{as Respondent R1 described:} ``\rev{[I]f it's a small shop \ellipsis and} they've developed all the software themselves, and they know it inside and out\ellipsis, \rev{the more likely it is that their `compliance,' such as it is, is `Hey guys,} anybody remember whether this has any third-party software in it?'\thinspace'' Larger companies engaged in government contract work, on the other hand, need to keep detailed records because they want to avoid delivering a product ``that [the government] can't use because it's got restricted software in it, or they wind up with, even worse, something that violates border control law where they're using software from overseas that [is prohibited].'' 
\looseness=-1

Even if the code's origin is clear, the nature of that origin can cause additional provenance problems, such as with snippets copy-pasted from other projects or pulled from coding sites like StackOverflow. Prior work has shown that snippets from such forums can be riddled with security vulnerabilities or bugs~\cite{acar2016you, fischer2017stack, ragkhitwetsagul2019toxic, tahir2018can} and lack attribution when reused~\cite{baltes2019usage, an2017stack}. \rev{Respondent R11} commented, ``[S]ometimes developers are doing exactly what we ask them to, which is to document where they're taking code from, [but] they ignore the fact that we say, `Please don't take code from Stack Overflow.'\ellipsis'' \rev{Respondent R15} offered a pragmatic perspective, stating \rev{that they tell developers}, ``If you pull a piece of code from some \rev{deep} Reddit thread, \ellipsis you don't know if [that user] legitimately acquired that code, \rev{developed that code, or anything like that\ellipsis.\rev{{[W]e need to have some degree of validation.}}}'' %
\looseness=-1

\vspace{-0.1cm}
\finding{Tracking software provenance is important for\\ compliance tasks, yet is still difficult in some situations.} %

\subsubsection{\textbf{Risk Management}}

Another important part of compliance is measuring and managing acceptable risk given the nature of open source enforcement. \rev{Respondent R14,} who favors greater enforcement, gave their take on the situation: ``[N]o one who's in business to make money is willing to comply with the license without a penalty, and they do a calculus. And they say, \ellipsis`What are the odds that we actually get caught?' And the odds are admittedly low.''  \rev{Respondent R15} told us that \rev{if a hypothetical client incorporated another party's functionality into its enterprise product without permission,} ``the truth is, because it comes out as a compiled product that is within an environment, the likelihood of the person that my theoretical client would have ripped off ever discovering that is slim to none.''  \rev{Other respondents conveyed} that it was very important to their organizations to be good ``open source citizens.'' %
As \rev{Respondent R2} put it, ``[W]ith every launch, there was inevitably some open source software developer that would reach out and say, `You're not complying with my license,' and then we'd fix it\ellipsis. I'm not sure most open source software developers are thinking of companies as targets. They're more concerned about `Are you a good citizen?' and `Are you trying?' \ellipsis \rev{[I]f you're blatantly just, `We don't care,' then I think people will come after you.}''
\looseness=-1

A company's risk tolerance may depend on where it operates. \rev{Respondent R15, who, as noted above, felt that the odds of being detected were low, stated} that ``the `move fast and break things' concept is one that American technology companies are very comfortable with\ellipsis. \rev{[Their mindset is that]} it's better to act and seek forgiveness than it is to seek permission\ellipsis; we'll just operate on this expectation, and if we need to buy somebody off on the back end, we will\ellipsis.  But in Europe, especially, and then in Japan, secondarily, they're much more risk averse when it comes to these things\ellipsis. When I have a conversation [with] my French clients, as opposed to my American clients, as opposed to my Japanese clients, \ellipsis I think of it as adjusting a mental dial before I get on the phone with them where I say, `Hey, this is what I think is an acceptable approach for this customer,' because there's a different risk tolerance across the board.'' %

Ultimately, however, Respondent R4 emphasized, "License compliance is absolutely important on the open source side, but I think it can also sometimes lead companies to view it as especially risky\ellipsis. [I]t's important not to overcorrect and treat open source as a kind of special beast. What's underlying all [software licenses] is that software is copyrighted, [and] licensors grant licenses in different ways."
\looseness=-1

\vspace{-0.1cm}
\finding{Risk management is essential during license compliance but may\\ depend on organizational culture and geographical location/jurisdiction.}

\subsubsection{\textbf{Education and Training}}

While there was an overall respect for software developers, many respondents noted that developers at times operate under false assumptions or incorrect information. \rev{Respondent R2} described \rev{a very senior developer} ``telling us how various licenses worked, and [there were] some kernels of truth in what he was saying, but some of it didn't actually line up with not just industry understandings but what the licenses said.'' \rev{Respondent R2 also noted} that developers' experience can give them greater \rev{(and sometimes misplaced)} confidence in their \rev{legal} understanding: ``Sometimes engineers, because of their experience or perhaps lack of experience in some cases, view themselves as quasi-lawyers\ellipsis. In many cases, they have a lot more experience in open source software than a lot of lawyers they're dealing with and might think, `I know how this works.' Often they don't.''  This state of affairs made \rev{Respondent R11} wonder what developers learn in school about best practices: ``Are they taught that you're just supposed to go out and grab whatever from the web and move fast and break things, \rev{or are you supposed to employ more diligence?}''
\looseness=-1

Given the lack of background in OSS licensing, training developers is critical for any (moderately sized) company; as \rev{Respondent R11} put it \rev{in their survey response}, ``The number one form of risk management is developer education.'' Training can, \rev{Respondent R13 noted,} ``arm [developers] to make good decisions early in the process. So by the time they [get] to us [the compliance review], they [have] a better chance of not having issues.'' \rev{Respondent R15} characterized the message as, ``You can't go out and create these [licensing] problems on the back end by saying, `Oh, yeah, I just downloaded this, and it seems to solve the problem. Therefore, I'm going to push it out to our customers.'\thinspace'' At a high level, the objective of training is, \rev{as Respondent R6 put it,} to ``convince the engineers that the compliance activities are going to be reasonable and possible to do and \ellipsis to convince the lawyers that the compliance activities are going to substantially reduce the risk.'' Training is not meant, however, to push the responsibility of compliance solely on developers. \rev{Respondent R13 stated that at the organizations they had worked for,} ``training was really focused on just providing background information to engineers \ellipsis so that they understood why we had the overall open source review and third-party software review process\ellipsis. It wasn't designed to have the engineers make the decisions on their own.'' %

The success of training efforts varies, with some developers internalizing content and others seeing it as a requirement to be tolerated, \rev{with the added challenge of employee turnover.} \rev{Respondent R11} described training as a constant process, like ``a game of whack-a-mole, where you're never catching up. I often describe my job as being kind of like in the movie \textit{Groundhog Day}, where I keep having the same conversations with different people.''  \rev{Respondent R13 characterized training as} ``effective in the sense that it brought general awareness and it got on people's radar that this was an issue and something they should be thoughtful about,'' \rev{while noting that it was not intended to give each employee} ``a detailed knowledge of the different licenses and [their] legal requirements.'' According to respondents, training sessions are typically recurring to ensure that the information stays fresh and are revised to reflect new developments, adapt to different learning preferences, or take account of areas where developers repeatedly have questions. In some companies, ``enhanced training'' exists for individuals who repeatedly or intentionally violate another party's intellectual property (IP).
\looseness=-1

Training is not the only way for legal practitioners to support development teams. Many also draft handbooks, licensing bibles, approval lists, or other forms of license clearance %
to offer engineers further guidance in their decision making \cite{riehle2019open}. (After training (19) and improved policies (9), licensing bibles (4) were the third most frequent educational method mentioned in the survey responses.) This guidance includes ``extensive wikis, FAQs, [and] internal documentation about licenses'' \rev{(R13),} as well as ``best practice documentations for our software team on things they should document as they're looking at open source technology'' \rev{(R16).}  Although guidelines for which licenses are permitted and under what circumstances are sometimes referred to as an approved/denied list, a red/yellow/green light system, or a license approval matrix, these methods are more nuanced than their names imply. As \rev{Respondent R13} stated, ``[I]t wasn't as simple as approved or denied licenses, but it was a spectrum, and we had pre-canned questions and advice based on what license and other parameters the engineers gave us.'' %

All of these methods are aimed at helping developers and teams make better decisions before getting legal counsel involved. \rev{Respondent R16 described their process:} ``I do have some green light, yellow light, red light stuff, clearly delineated on those lists, so [the developers] understand \ellipsis when they need to pick their head up and have a proactive conversation with me before they start to incorporate technology.''  \rev{Such} resources support a ``self-service model [of] education, so that I don't have to be integrated in the entire process and that [the engineers] can also educate their colleagues as they move forward.'' %

\finding{Developers may have an incomplete or incorrect understanding of compliance requirements, making training essential. Since it can be difficult to get developers invested in training, lawyers can create additional resources (\eg guides and bibles) that help developers with compliance tasks.}

\subsubsection{\textbf{Tooling}}

Tool usage is often employed in compliance tasks, particularly to detect and track third-party components. Of the 30 respondents, 21 indicated that they (and developers) used tooling.  The most popular offerings were Black Duck~\cite{blackduck} (11), FOSSology~\cite{fossology} (4), and ScanCode~\cite{scancode} (3). As \rev{Respondent R6} said, ``These days, you really can't do [compliance] without tooling\ellipsis. [M]ost sophisticated products have way too many components for anyone to keep track of them by hand.'' %

Respondents reported that tools are primarily used during the development process or prior to shipping software to validate license compliance. \rev{Code scans can also be used to determine, as part of a risk mitigation strategy, the code that a vendor or an acquisition target is using.}
  \rev{Respondent R7 suggested that, to the extent that vendors can provide tools that scan binaries,} organizations and developers can also \rev{use such tools} to \rev{identify their open source code in a proprietary software product and} ``use that as a mechanism to tell [the distributor] that they're not complying with their license obligations.''
\looseness=-1

Through our interviews, we distill three main steps in the compliance process: 1) identifying third-party components being included, 2) determining the licenses of those components and how those components will be used, and 3) applying the client's compliance policies to that information. \rev{Respondent R4} assessed that tooling is ``especially essential for step one, partly essential for step two, and not really that relevant to step three.''  Similarly, \rev{Respondent R11} stated, ``I trust the computer's ability to match code better than I trust the human [who] input the license information into the database. And so I almost disregard the license information, and I just really [ask], `Did [the tool] find something?'\thinspace''
\looseness=-1

A few respondents described the current state of tooling as ``pretty good.'' \rev{Respondent R8} was ``not aware of too many situations where clients have used a tool like that and it missed something that was open source.''  \rev{Respondent R7} believed that clients use tools because they ``are pretty good in terms of having massive databases of open source code against which they compare\ellipsis. \rev{[It's] not certain that you'll catch everything, but you're probably going to catch a good percentage of stuff, enough that you can have} a high degree of confidence that you've identified at least third-party open source code that's part of your stack.'' %
\looseness=-1
  
Nevertheless, tools are still imperfect. One of the primary limitations mentioned in our study was the accuracy of the tools. \rev{Respondent R11 said, referencing the licensing information in the database,} ``Probably more often than not, it's accurate. But when it's inaccurate, it's really painful,''\rev{ which can lead to problems when users ``place a lot of trust in that rather than going and trying to verify it for themselves.''}  \rev{Respondent R6} noted that in their experience, tools ``usually don't capture [meta information] properly,'' which is why they ``never rely on the tools for that information because \ellipsis most of the time it's inaccurate.'' %

One of the main inaccuracies mentioned in our discussions was false positives, which can cost a team or organization valuable time. Respondents discussed cases where tools were reporting as a match ``things that wouldn't be eligible for copyright protection anyway, \ellipsis  like class names'' \rev{(R11).} Tool output inaccuracies may result from how the tools are configured. As \rev{Respondent R13} put it, some tools have a setting ``where you can search for key terms that you find problematic, and that can be really noisy depending on how you configure it.''  \rev{Respondent R2} reasoned that tools are ``just being overly cautious,'' which leads them to ``identify a lot of false positives, things that they're saying [are] in the code base, and then when you actually dig into it with your software engineers, those pieces are not in the code base.'' Another type of false positive, \rev{identified by Respondent R5, is one that} presents itself as a license problem ``when really it's just a missing word or misidentified license.'' In all of these cases, determining if a detection is actually a problem ``requires a much deeper dive'' \rev{(R5)}. These false positives can result in a financial loss because, \rev{as Respondent R2 noted,} organizations ``have to pull these very busy developers off all the things they're doing and say, `Hey, let's go through this list.' It's the last thing on earth they want to do.'' %

\rev{Respondents reported tools' limited} ability to provide analysis of compliance issues. Tools are generally good at providing information, but human review is required to know what to do with that information. \rev{Respondent R6} described it as ``quite frustrating, because the output of the tools is information but not answers,'' and \rev{Respondent R1} said the current capabilities are ``a far cry from human knowledge.'' Additionally, \rev{Respondent R13 noted,} ``\rev{it's really hard for the tools to make decisions about whether something actually complies with the license, because} most of the tools don't have awareness of how the code is integrated and used within your proprietary code.'' \rev{As Respondent R4 put it,} the ``tendency for automated tools to do a lot of red, yellow, green categorizing \ellipsis  misses a lot of things and can lead clients in the wrong direction on either end of the spectrum.'' Respondent R9 even reported that tools seem ``to ignore important parts of the license and focus on parts that are still important but not the ones that we're seeing most commonly violated.'' %

As described in \Cref{ssec:actors}, there was a general consensus among respondents that developers should be the primary tool users. \rev{Respondent R6} said, ``The tools absolutely need to be for developers. \rev{When compliance is done correctly, it is 95 percent an engineering decision. There are very small numbers of cases where lawyers really should be brought in, but mostly, you have a set of rules, and the engineers comply with those rules, and they should never have to get the lawyers involved because that will just slow down their process.''} Nevertheless, respondents indicated that any user of a tool needs to be aware of its limitations. \rev{Respondent R4} noted the importance of tool developers being ``clear with their users \ellipsis [who need to] have the understanding that none of these are a silver bullet [and that] these are all solving different pieces of a much bigger question.'' Tools can also be difficult to set up and use; \rev{as Respondent R13 noted about some tools:} ``\rev{They're} not tools [where] you can simply deploy them and they run themselves, \rev{especially at companies like the size that I supported}. [T]hey require dedicated teams to manage the outputs and triage the results.'' %

Interviewees suggested several improvements to tooling.
\rev{Respondent R11 provided some examples. First,} ``[b]etter maintenance of the upstream location where the tool is pulling information from. A lot of times there'll be dead homepages. It's not kept up to date. They'll put something into the catalog, and then a couple of years go by, and it's moved location, and it takes some human engineering to figure out where the code currently is or what the status currently is.''  \rev{Second,} tools that ``produc[e] a really good notice file, not just like a phone book\textendash sized SPDX dump, but actually something that's human readable and useful.''  Lastly, tools ``that can do both license and security checking at the same time. It would be beautiful if we only had to scan a code base one time with one tool.'' \rev{Respondent R7 offered a similar suggestion:} ``The one thing that I think would be of great use is one tool to rule them all, and one tool to rule them all that doesn't cost an arm and a leg\ellipsis.'' %
\vspace{-0.1cm}
\finding{Organizations commonly use tools to identify software components and their licenses. \\However, tools provide little to no explanation and analysis of licensing issues. \\The number of false positives must be reduced for tools to be more effective and practical.}

\subsection{RQ3: Challenges that Legal Practitioners Face during License Compliance}

Respondents described a number of challenges that legal practitioners face during license compliance. We focus here on
changes in the way that software is developed and distributed; interpretation of license language; and communication between the legal team and the engineering team.
\looseness=-1

\subsubsection{\textbf{Changes in the Way Software is Developed and Distributed.}}
The way software systems are integrated and built has changed over the years, with more and more components being integrated into new systems.
The number of components and dependencies typically involved in even a small project has a direct effect on a lawyer's compliance work. Each component must be verified to ensure that its license is being complied with, \rev{which ultimately} requires human review. \rev{Respondent R15} recounted an experience in which the client went through a significant acquisition, which involved running a scan of all involved components; the resulting list of components alone was 50 pages long.  \rev{Respondent R15} concluded, ``It is honestly impossible to do an effective review of that. We could evaluate and we could say, `OK, fine. These things are important; these things aren't.' But just as a practical matter, there is no useful tool that will determine whether or not something infringes.'' \rev{Respondent R6} conveyed that  licenses can create ``an enormous administrative burden \rev{on people using software}. Don't get me wrong — it was perfectly reasonable to do, particularly at the time open source licensing started, but I'm not sure that even the most ardent open source or free software advocate in the 1990s contemplated that people would be selling products that had thousands and thousands of components.'' This has the potential to pose a particular burden on smaller companies, \rev{as Respondent R6 also noted}: 
``[M]ost companies care about being compliant, but, honestly, very small companies are just flying by the seat of their pants and their compliance is all just in time\ellipsis. [T]hey just can't afford compliance activities or compliance tools.''  %

At the same time that software development has become more complex, the methods of distributing information have become easier, a fact that that the licenses do not always incorporate.
As \rev{Respondent R6} \rev{explained}, ``\rev{[Licenses] require you to deliver full-text copies of the license and so forth\ellipsis.} Clients ask me all the time, `Why do I have to do this? Does anyone look at these? Can't I just post them on a website?' And the answer to that question should be yes, except for [the fact that] GPL 2 was written in 1991\ellipsis. \rev{But open source license drafting at this point is paralyzed because of divisions in the community\ellipsis. I don't think the right people have a desire to fix this problem.''} 
\looseness=-1

\finding{Software development and systems have become more complex over time due to the increasing number of components/dependencies, making lawyers' compliance efforts harder.}

\subsubsection{\textbf{Interpretation of License Language.}}
OSS licenses are documents intended to have legal effect, with the potential to be enforced in court. As with other legal documents, parties may have different interpretations of ambiguous terms in an OSS license. As \rev{Respondent R7} noted, ``[T]he law doesn't work like a computer does, [where] you put inputs in, and you know what the output is.'' %
\looseness=-1

This interpretive difficulty arises in part, some %
interview respondents suggested, because most OSS licenses were drafted and promulgated by software engineers and others in the SE field and not by lawyers. Thus, as one survey respondent, \rev{a lawyer in private practice,} noted, such licenses ``sometimes are written using terms that don't directly map onto legal doctrine or are just ambiguous (perhaps intentionally), and questions then arise [as to] how best to understand them.'' This causes what \rev{Respondent R13} stated to be ``one of the downfalls of open source licenses'' — ``the text is static in time, but obviously technology is moving,'' sometimes resulting in a situation where ``[you] don't think that the authors intended the license to have this impact, but it does just because the wrong terms were used given the new technology.'' %
\looseness=-1

Two examples of interpretive uncertainty were mentioned by interview respondents. %
First, the requirements of GPL v2 apply to modifications that form a work ``based on the Program,'' which GPL v2 defines as ``any derivative work under copyright law: that is to say, a work containing the Program or a portion of it, either verbatim or with modifications and/or translated into another language.'' The question then arises whether, for example, an application under GPL v2 that makes a system call to a library creates a unified work ``based on the Program'' or whether the library remains a distinct program. The GPL FAQs (Frequently Asked Questions) state~\cite{gnu_faq}, ``Linking a GPL covered work statically or dynamically with other modules is making a combined work based on the GPL covered work. Thus, the terms and conditions of the GNU General Public License cover the whole combination.'' But U.S. copyright law is not as clear on this question, as cases such as \textit{Lewis Galoob Toys, Inc. v. Nintendo of America, Inc.}~\cite{Lewis} illustrate. 
\looseness=-1

The second example involves when an activity includes distribution, which is the trigger for compliance in several licenses. Interview respondents who raised this issue noted distinctions between products that are provided to end users versus those used only internally (which are arguably not distributed), although SaaS~\cite{cusumano2010cloud} complicates this question. \rev{Respondent R8} gave another scenario: ``[GPL v2] says [that] if you distribute software that includes an GPL component, then your distribution has to be covered under GPL. But if you distribute something that works with something else [that] gets called at runtime but it's not part of your code, and there's a dynamic link or some other kind of communication between the programs, to me, you're not distributing that GPL program\ellipsis. So there's issues like that [one] that are pretty significant.'' %

The presence of ambiguous terms is not a fatal obstacle to enforcement; it is rare that a legal document is not open to interpretation in some respect. But respondents reported that because relatively few disputes in the OSS licensing space are resolved through litigation, there is a paucity of caselaw from U.S. courts on how OSS license terms should be interpreted, compounded by the existence of many unresolved issues in U.S. copyright law generally. This means that lawyers look to community resources and norms as a guide to license interpretation. These norms are particularly relevant because, as \rev{Respondent R2} described it, ``it's less about legal sometimes and more about just playing nice with the community. That dynamic is very, very important.'' %

FAQs and similar publications developed by license stewards were reported by respondents to be a primary source of authority on OSS license interpretation. %
FAQs by the Free Software  Foundation~\cite{gnu_faq} %
were seen as particularly influential, even though, in \rev{Respondent R6's} view, they are ``long, complicated, and old.'' %
In addition to FAQs, interview respondents mentioned other sources of community beliefs of license meaning: listservs, mailing lists, and other community channels; sources such as Hacker News~\cite{hacker_news} (a social news website run by the startup incubator Y Combinator); and information posted on GitHub and other documentation sites by developers.

Although the community might look to the views of license stewards as to the meaning of the licenses they drafted, it is unclear whether a court would also do so. %
\rev{Respondent R10} put it this way: ``If you're just cutting and pasting a license that the Free Software Foundation wrote or Mozilla wrote, who's the drafter? Whose intent actually matters at that point? Because, sure, the Free Software Foundation is the author of that license, but they're not the licensor\ellipsis. Are you [now] the drafter because you chose what to cut and paste? Does your intent as the licensor matter? That becomes a really interesting question.'' %

As a practical matter, the lack of interpretive guidance from courts, and the shift in interpretative authority to the community, affects the nature of the advice lawyers give to their clients. In some cases, lawyers have to explain to clients that the community's view may not be cognizable in court; in other cases, compliance becomes a matter of conforming to community norms rather than conforming to legal requirements. \rev{Respondent R6} stated, ``[W]hen a client comes to me and says, `What I should do?,' I advise them what to do in order to meet community practice. I can't really advise them what to do to meet the letter of the law, because the letter of the law is too vague.'' %
Additionally, it is difficult to know in advance whether a particular norm will apply equally to all participants. As \rev{Respondent R13} stated, ``[M]aybe a competitor can do X, Y, and Z, [but] we might not be treated the same. So we would have to factor that into our analysis.''  Similarly, since there is no single arbiter of interpretive uncertainty, whether the ``correct'' answer has been reached in any particular case is left up in the air. \rev{Respondent R15} characterized it as follows: ``[C]ertainly I and my clients set out with the goals of conforming to open source licenses. But even with our best efforts, we can't be sure that we are honoring either the intent or the letter of what the licensor is trying to do\ellipsis. [I]f you look at some of the open source communities and the sources for some of the licenses, they will offer their gloss as to what they think [the language] means. But \ellipsis  absent a court's interpretation of how that language would be honored by a court in the case of litigation, we're all just kind of guessing\ellipsis.'' %

As a result, lawyers might, as a form of risk management, advise clients to steer clear of any interpretive gray areas. \rev{Respondent R10} stated, ``My experience talking to other attorneys in the space has been that, for the most part, no one really wants to push the limits too much\ellipsis. [T]o the extent the ambiguities exist, they kind of want to \rev{stay within [the lines] enough that [the ambiguity] doesn't need to be clarified.}'' \rev{Respondent R13} noted that although the client might have a strong legal argument, ``maybe we don't want to be the ones defending that argument\ellipsis. [It's] easier and simpler to just align with the community's view on the topic than to get into a discussion and try to walk in that gray area.'' On the other hand, \rev{Respondent R13} acknowledged, there are times when the client needs to take a stand: ``[It's] hearing people out, trying to educate, trying to understand, `Is there a middle ground?' But some cases aren't just as simple as, `Oh, we'll just fix it and move on.' Sometimes it is fundamental to the architecture. We just fundamentally disagree.''
\looseness=-1

To be sure, not every interview respondent shared a similar view on the interpretive difficulties of OSS licenses. \rev{Respondent R9,} who is involved in OSS license compliance but who is not trained as a lawyer, %
conveyed the view that the obligations under many OSS licenses are clear and that entities that claim interpretive uncertainty are trying to ``figure out what the rest of the industry is doing'' when they should instead ``make a conservative interpretation'' of ambiguous language.  \rev{Respondent R10} stated that in transactional work, ambiguities ``don't come up that often'' because ``there is a pretty good consensus among free and open source software attorneys, at least in the United States, about how these licenses work in the vast majority of cases.'' \rev{Respondent R14,} who favors strong enforcement of OSS licenses, characterized interpretive debates as ``a political struggle between firms who want certain licensing outcomes \ellipsis that benefit their businesses'' and those who ``care about the rights and freedoms of users who receive software.'' %

Respondents generally seemed to be aware of this diversity of views, the result of the fact that the OSS legal community is relatively small. Indeed, \rev{Respondent R10} expressed concern about whether things would change as more commercial entities ``who don't have an interest in the license''  incorporate OSS into their offerings. \rev{Respondent R13} noted, similarly, ``The community takes a different view on some topics than a large software company,'' highlighting, in this phrasing, that large software companies are, for some, not seen as part of ``the community.'' Lawyers in this field can therefore gain their own interpretive authority, as \rev{Respondent R14} stated, ``merely by being around for a long time.'' %

\finding{OSS licenses pose interpretive challenges because licenses are static while technology evolves over time, licenses were drafted by SE/CS experts rather than lawyers, and there is a lack of interpretive guidance from U.S. courts. As a result, lawyers rely on community norms and best practices (documented in sources such as FAQs and mailing lists) to interpret licenses and provide legal guidance to their clients. Lawyers sometimes advise clients to avoid ``grey areas'' due to interpretive uncertainty.}

\subsubsection{\textbf{Communication Between the Legal and Engineering Teams.}}
Effective OSS license compliance ultimately depends on productive relationships between lawyers and developers. %
Legal professionals with a background in SE or a related field believed that this experience helped to facilitate communications with engineers because it gave them a certain level of credibility when discussing SE issues. \rev{Respondent R10} reported that they would use this background to write ``toy programs'' to illustrate to software engineers the licensing impact of certain development decisions because ``if you can't talk to those engineers, if they don't understand you, [if] they don't have that respect, your advice really isn't going to go anywhere.''  By contrast, \rev{Respondent R1,} who does not have a software engineering background, described using an intermediary to facilitate communications with software engineers employed by Respondent R1's clients. %

A particular challenge reported by respondents was a need for developers to understand the value contributed by legal counsel. One survey respondent, who is not trained as a lawyer but consults in the compliance arena, stated that ``the worst'' situation ``is when there is a local `tech guru' who thinks [they know] how the licenses work but [who] has never actually read the actual license texts, and everyone in the company goes to that guru. It is super counterproductive.''  A related challenge arises from cultural differences in attitudes toward intellectual property that originate in a developer's past experience. As \rev{Respondent R11} stated, ``Sometimes \ellipsis getting [people] to appreciate the value of IP and the importance of respecting other people's IP can be an interesting opportunity or a challenge. They don't see anything wrong with [what they're doing], or they've been like, `Well, I've been developing all this time doing this thing.' And it's like, `[W]ell, we don't do that here.'\thinspace'' %
\looseness=-1

\looseness=-1 Ultimately, interview respondents reported a need to ensure that developers and lawyers know that they are on the same team, both in terms of compliance and in terms of the shared business interest in putting out a good product, \rev{whether or not it involves OSS}. As \rev{Respondent R2} put it, ``the key to being a good lawyer is becoming a \ellipsis partner rather than a police officer. You are there to enable them to get their work done. Sometimes you have to say no just based on policy, but other times, you've got to try to use creativity to help them get to where they're trying to get.'' \rev{Respondent R14,} who is not trained as a lawyer but who has expertise in compliance issues, expressed skepticism in their survey response about this relationship, stating, ``In my experience, what I've seen is organizations tend to be told by their legal departments to `just trust us' that we know how to comply with the license.'' \rev{Respondent R3} offered a similar thought, noting, ``I think for a lot of companies, once they get a lawyer involved, they tend to outsource a lot to the lawyer and just assume there's standard forms and clauses that everybody just accepts.'' \rev{Respondent R4}, however, offered this perspective \rev{in their survey response}: ``The legal teams establish policy and provid[e] guidance, while the engineering teams have to take that guidance and apply it to particular technical use cases. Without trust and open communication between those teams, it is impossible to resolve [licensing compliance] challenges.'' %

\finding{Communication and trust between lawyers and developers can be challenging, in part, due to a lack of developer understanding of the value of legal advice, different cultural attitudes toward intellectual property, and different company norms about how lawyers and developers should interact.}

\section{Threats to Validity} 
\looseness=-1
\textbf{External Validity:} %
The conclusions drawn in this study apply only to the population that participated in our survey and follow-up interviews.  We cannot generalize our results, but in light of the themes that emerged, we believe that other lawyers in the OSS community will share some of the same experiences. That said, our goal was not to claim generalizability but to attain a fuller understanding of the software compliance landscape from a legal practitioner's perspective and to identify current practices and challenges.

\textbf{Internal Validity:}
To mitigate bias, we used an open coding methodology for both the survey and interview transcripts and had SE and legal researchers involved in every stage of the process.  We employed diverse strategies to locate participants (professional networks, mailing lists, top law firms, \etc) to increase the pool of different perspectives and minimize potential bias, but we are aware of the issues that may arise from low response rates and self-selection bias.  We followed best practices in the formulation of survey and interview questions, making sure that questions were written clearly and concisely to avoid confusion and biasing language. Interviews were limited to at most an hour, meaning that not all of a participant's views were likely heard. We limited confirmation bias in qualitative analysis by independent coding, discussing disagreement, and arriving at a consensus based on the data.

\section{Related Work}

\textbf{License Compliance, Practices, and Needs.}
Given the challenges and importance of license compliance, researchers have developed a number of tools and processes to assist with license compliance tasks (\eg identifying licenses, detecting and fixing license incompatibilities)~\cite{tuunanen2009automated, ombredanne2020free, kapitsaki2017automating, german2009license, german2012method, german2010understanding,hemel2011finding, feng2019open}.
Other works catalog and detect non-approved licenses, license variants, exceptions, and questions by mining software repositories \cite{meloca2018understanding, zacchiroli2022large, papoutsoglou2022analysis, vendome2017machine}.
Prior work also explored when, why, and how developers change their software's licenses \cite{di2010exploratory, vendome2017license}, as well as ways to predict license changes \cite{liu2019predicting}. %
\rev{Developers have been a primary focus of previous investigations, which have demonstrated that they} tend to have difficulty understanding OSS licenses \cite{almeida2017software, almeida2019investigating} \rev{as well as identified the types of questions they have about software licenses ~\cite{kapitsaki2020developers}}. 
Other work addresses this by investigating the automatic summarization of legal documents \cite{manor2019plain} and proposing systems of recommending licenses \cite{kapitsaki2019modeling, liu2021choosing}. \rev{While some recent works have proposed using formalization and rule-based decision making to assist with cases where legal interpretation may be unambiguous ~\cite{merigoux2021catala, verheij2017formalizing}, such automated systems introduce concerns that have yet to be addressed, including determining how to handle ambiguities in language, reduced transparency of the legal process, and questions regarding who would be responsible for errors ~\cite{guitton2023mapping}. Gangadharan \etal ~\cite{gangadharan2012managing} apply formalization to software licenses, but solely to address license compatibility.}
All this prior work does not examine the state of the practice of \rev{software} license compliance from \rev{the perspective of legal practitioners}, as we do. 

\textbf{Licensing Bugs and Violations.}
\rev{Licensing bugs and violations can be introduced into software through a variety of different methods, such as via code clones from other software ~\cite{german2009code} or Q\&A websites ~\cite{an2017stack, baltes2019usage}.} Prior work cataloged licensing bugs and incompatibilities based on analysis of OSS project repositories~\cite{vendome2018distribute,wu2015method,wu2017analysis,golubev2020study}.
Several studies also indicate that licensing bugs are prevalent in modern software ecosystems, such as the NPM, RubyGems \cite{makari2022prevalence,meloca2018understanding}, Android~\cite{mlouki2016detection}, PyPI~\cite{xu2023understanding}, and JavaScript \cite{qiu2021empirical} ecosystems. Moraes \etal~\cite{moraes2021one} also show the prevalence of multi-licensing in JavaScript, further complicating the licensing landscape. 
All this prior work has focused on repository mining, while we conduct surveys/interviews with legal practitioners.

\section{Implications and Conclusions}
\label{sec:implications}

By qualitatively analyzing the survey and interview responses of 30 lawyers and other individuals who specialize in OSS license compliance, we have identified: (1) the state of the ecosystem of OSS license compliance, (2) how legal practitioners perform compliance, and (3) the challenges faced by legal practitioners during compliance. Our findings warrant further research intended to support 
developers, legal practitioners, and other roles in performing license compliance more effectively, which is essential for software engineering companies and the OSS community. We now discuss the implications of our findings.

\textbf{Robust tooling is needed.} %
Compliance has become more difficult at the scale of modern software, where a single product might contain thousands of components. Given the enormous scale of compliance tasks today, manual compliance analysis is likely infeasible for large software products. As such, the importance of automated tools to assist with license compliance only increases. Yet, as our participants identified, current tooling can be difficult to use, is prone to false positives, and is limited to providing data rather than analysis. This demonstrates a clear need for more robust license compliance tools that are accessible to both developers and legal practitioners, keeping in mind the inherent analytical limitations of any tool.

\textbf{Effective communication and interaction between lawyers and developers is needed.}
The optimal process is not for developers to build a software product and then pass it on to lawyers for compliance. Instead, compliance-related conversations should take place throughout the process. This requires effective communication between the teams to ensure an understanding of each team's domain. Legal professionals can facilitate this by creating educational and actionable resources for developers so that they can be aware of the legal implications of development decisions and can more independently make decisions that will not negatively impact the organization at large. Indeed, our study suggests that similar educational efforts should take place early in developers' careers -- during their undergraduate and graduate training -- so that legal compliance is understood to be an integral part of software engineering, not separate from it.

\textbf{Integrated and continuous license compliance.}
We saw in our conversations that proactive approaches to compliance that are done early and often lead to the best outcomes. Put another way, from the inception of a project, \textit{compliance should be treated as a nonfunctional requirement}. 

Compliance tasks should start before the first lines of code are written and should be regularly assessed throughout the development process. For example, the compliance obligations of third-party components should be vetted \textit{before} incorporating them into a build. Regular scans can also verify that no dependencies were slipped in under the radar. Tool-based license compliance checking can be incorporated into continuous integration/delivery pipelines to detect potential licensing issues in the process. Additionally,  well-understood agile development practices can also be easily adapted to facilitate license compliance tasks. For example, if records are kept during sprint meetings, it can be much easier to track engineering decisions and software provenance. 

It may take time for developers to adjust to a more compliance-minded approach, as some software developers, particularly those who work in small teams or on personal projects, have the mentality of ``move fast and break things.'' Future work can explore the information needs and requirements for an \textit{integrated and continuous compliance process} that works in tandem with 
development processes.
\looseness=-1

\textbf{Community norms and license enforcement.}
Software has changed significantly since many OSS licenses were first drafted. There appears to be little interest, however, in bespoke licenses with updated terms; legal practitioners strongly prefer using an existing license. In addition, the use of community norms as a source of interpretive guidance means that ambiguities will persist. OSS license interpretation will therefore be a significant challenge for the foreseeable future, which increases the likelihood of (perceived) noncompliance. Going forward, the risk of interpretive gray areas can be addressed in several ways, including attention to known ambiguities by the drafters of future license revisions; continuing the trend away from license proliferation; and greater collaboration and convergence on published sources of interpretive guidance.

\section*{Data Availability}
We provide an anonymized replication package containing survey and interview protocols, aggregated results, a code catalog for survey and interview responses with definitions, code to process results, and other data required for verifiability~\cite{anonymous_repo}.

\section*{Acknowledgments}
We would like to acknowledge Sang Hwan Lee and JT Shorten for their contributions to this project.  Without their help in drafting the initial survey and reviewing responses, this work would have been impossible. We also thank the study participants for their time and valuable contributions. This research has been supported in part by NSF grant CCF-2217733. Any opinions, findings, and conclusions expressed herein are the authors’ and do not necessarily reflect those of the sponsors. A complete, detailed list of image attributions can be found at \cite{anonymous_repo}.

\balance

\bibliographystyle{ACM-Reference-Format} %
\bibliography{references}

\end{document}